\documentclass[reprint,pra]{revtex4}
\usepackage[T1]{fontenc}
\usepackage[latin9]{inputenc}
\usepackage{color}
\usepackage{amsmath}
\usepackage{amssymb}
\usepackage{graphicx}
\usepackage{esint}
\usepackage{bm}
\usepackage[english]{babel}
\usepackage{appendix}
\usepackage{braket}
\usepackage{float}
\usepackage{cleveref}

\newcommand{\Tr}{\operatorname{Tr}}

\makeatletter \@ifundefined{textcolor}{} {
\definecolor{BLACK}{gray}{0}
 \definecolor{WHITE}{gray}{1}
 \definecolor{RED}{rgb}{1,0,0}
 \definecolor{GREEN}{rgb}{0,1,0}
 \definecolor{BLUE}{rgb}{0,0,1}
 \definecolor{CYAN}{cmyk}{1,0,0,0}
 \definecolor{MAGENTA}{cmyk}{0,1,0,0}
 \definecolor{YELLOW}{cmyk}{0,0,1,0}
 }

 \newtheorem{theorem}{Theorem}[section]
 \newtheorem{defn}{Definition}[section]

\begin{document}

\title{Reply to ``Comment on `Gleason-type theorem including qubits and projective measurements: the Born rule beyond quantum physics' '', by Michael J. W. Hall}
\author{F. De Zela}

\affiliation{ Departamento de Ciencias, Secci\'{o}n
F\'{i}sica, Pontificia Universidad Cat\'{o}lica del Per\'{u}, Apartado 1761, Lima, Peru.}

\begin{abstract}
We present a reply to the objections raised by M. J. W. Hall against our extension of Gleason's theorem.
\end{abstract}

\maketitle

\section{Introduction}

In a recent Comment \cite{Hall}, Hall argues that the extension of Gleason's theorem \cite{Gleason} that was proposed in Ref.~\cite{fdz} is flawed. Hall presents a counterexample to the main result in Ref.~\cite{fdz} and briefly discusses an alleged flaw in the derivation of the proposed extension of Gleason's theorem. Hall's Comment represents a welcome opportunity to expand somewhat the scope of Ref.~\cite{fdz}, as well as to clear up the physical content of the proposed extension of Gleason's theorem. As we shall see, there is no technical flaw in this extension.

\section{Gleason's theorem and its extension}

As is well known, Gleason's theorem does not apply to qubits. Hall \cite{Hall} gives a simple counterexample that illustrates why this is so. The counterexample consists of a positive-definite function $p_s(P_{\psi})$ on the lattice of qubit projections $P_{\psi}$ that, while satisfying Gleason's requirements for a probability measure, is a non-linear function. This latter feature precludes that $p_s(P_{\psi})$ coincides with Born's rule, i.e., with the quantum-mechanical linear expression $\Tr(\rho_s P_{\psi})$ for a probability measure. A similar argument is used by Hall for the extension of Gleason's theorem that was proposed in Ref. \cite{fdz}. As for the alleged flaw in \cite{fdz}, Hall argues that it derives from a misuse of Gudder's theorem \cite{Gudder}. This theorem deals with an inner product vector space $V$ and a continuous function $f$ that is orthogonally additive. The definition of such a function reads
\begin{defn}\label{def1}
\begin{equation}
\text{$f:V \rightarrow \mathbb{R}$ is orthogonally additive, if} \quad
  f(r+r^{\prime})=f(r)+f(r^{\prime}) \quad \text{    whenever  } \quad r \cdot r^{\prime}=0.
\end{equation}
\end{defn}
Gudder proves that the following result holds true:
\begin{theorem}\label{t1}
\text{If $f:V \rightarrow \mathbb{R}$ is orthogonally additive and continuous, then it has the form} \quad
\begin{equation}
  f(r)=c (r \cdot r)+ k\cdot r,
\end{equation}
where $c \in \mathbb{R}$ and $k\in V$.
\end{theorem}

Our aim is to show how Born's rule arises from some fundamental physical notions. To this end, we resort to Gudder's theorem. A very basic concept in physics is that of a ``measure''. This is a concept that underlies most experimental procedures in physics. The latter are essentially ``counting'' procedures, i.e., procedures that consist in counting how many times a given unit fits into an observable that is submitted to measurement. As for the mathematical tool that corresponds to our basic notion of a measure, it is defined as a non-negative application $m$ over a $\sigma$-algebra, which
is required to satisfy $m(A \cup B)=m(A)+m(B)$, whenever $A\cap B =\emptyset$. Last condition must hold because in case $A\cap B \neq\emptyset$, we should subtract $m(A\cap B)$ from $m(A)+m(B)$ in order to encompass our intuitive notion of a measure. A particular measure is
the ``probability measure''. In quantum mechanics, this measure is defined over the projection lattice
$\mathcal{P}(\mathcal{H})$, and it is thus consistent to require for $P_i,P_j \in \mathcal{P}(\mathcal{H})$ that $m(P_i+ P_j)=m(P_i)+m(P_j)$, whenever $P_i P_j=0$. On the other hand, it is rather
unnatural to call $v$ a measure, if it is required to satisfy
$v(E_n +E_m)=v(E_n)+v(E_m)$ even though $E_n E_m \neq 0$. This is however the case in Busch's extension \cite{Busch} of Gleason's theorem, in which projectors $P_i$ are replaced by positive operator-valued measures $E_n$.

We can expect that an orthogonally additive function, cf. Definition \ref{def1}, can give rise to a measure. In such a case, besides the additivity property that is required from a measure, we also require that $f$ is continuous. This last requirement captures our basic notion that infinitesimal variations of the observable being measured should lead to infinitesimal variations of the measurement result. On the other hand, we do not yet assume that $f\geq 0$, as it would be required for $f$ to be a measure.

Let us now focus on qubits.
A qubit can be represented by a unit
vector $|\phi\rangle \in \mathcal{H}_{2}$ of an equivalence class, a so-called ``ray'', or --alternatively -- it can be represented by the
corresponding projector
\begin{equation}
P_{\phi }\equiv| \phi \rangle \langle \phi | =\frac{1}{2} \left(
\openone+\mathbf{\hat{n}}_{\phi }\cdot \boldsymbol{\sigma }\right) .
\label{proy}
\end{equation}
Here, $\openone$ is the identity operator in $\mathcal{H}_{2}$ and the unit vector $\mathbf{\hat{n}}_{\phi }=\Tr \left(%
\boldsymbol{\sigma }P_{\phi }\right)$, with $\boldsymbol{\sigma }$
the triple of Pauli matrices.
In general, for a
non-normalized qubit $|\psi\rangle \in \mathcal{H}_{2}$ we can write
\begin{equation}
R_{\psi }\equiv| \psi \rangle \langle \psi | =\frac{1}{2}
\sum_{\mu=0}^{3} r_{\mu} \sigma_{\mu}, \label{roy}
\end{equation}
with $\sigma_{0}\equiv \openone$ and $r_{\mu}=\Tr \left(%
\sigma_{\mu} R_{\psi }\right)$. We see that $R_{\psi }=P_{\psi }$,
whenever $\langle \psi|\psi\rangle =1$. There is a one-to-one
correspondence between operators $R_{\psi }$ and vectors
$r:=\left( r_0,r_1,r_2,r_3 \right)\equiv (r_0,\mathbf{r})$. The
latter span a four-dimensional real vector space $V_{4}$ that can
be made an inner product space by defining the Euclidean
inner product
\begin{equation}
r \cdot r^{\prime}=\sum_{\mu=0}^{3} r_{\mu}r_{\mu}^{\prime}.
\end{equation}

We wish now to define a measure $f_{\phi}$ that is associated to a particular qubit
$|\phi\rangle \leftrightarrow r_{\phi}\equiv (1,\mathbf{\hat{n}_{\phi}})$. In a sense, $f_{\phi}$ and
$|\phi\rangle $ represent one and the same physical object \cite{fdz}. To start with, $f_{\phi}$ must satisfy the requirements of Theorem \ref{t1}. Furthermore, we naturally require that $f_{\phi}(r_{\phi })=1$, which corresponds to requiring that our unit of measure fits exactly one time into itself. We also naturally require that for the qubit $|\phi _{\perp}\rangle \leftrightarrow r_{\phi_{\perp}}\equiv (1,-\mathbf{\hat{n}_{\phi}})$ that is orthogonal to $|\phi \rangle$, it holds $f_{\phi}(r_{\phi _{\perp}})=0$.
We have then, on applying Gudder's theorem with $k=(k_0,\mathbf{k})$,
\begin{subequations}
\label{mf}
\begin{eqnarray}
f_{\phi}\left[(1,\mathbf{\hat{n}_{\phi}})\right]&=& 2c+k_{0}+\mathbf{\hat{n}_{\phi}}\cdot \mathbf{k} = 1, \label{mf1}\\
f_{\phi}\left[(1,-\mathbf{\hat{n}_{\phi}})\right]&=& 2c+k_{0}-\mathbf{\hat{n}_{\phi}}\cdot \mathbf{k} = 0. \label{mf2}
\end{eqnarray}
\end{subequations}
From these equations we get $2c+k_0=1/2$ and $\mathbf{\hat{n}_{\phi}}\cdot \mathbf{k} = 1/2$.
Up to this point, we are dealing with a function $f_{\phi}$ that is not necessarily identifiable with a probability measure. Let us further restrict $f_{\phi}$ to satisfy the requirement
$f_{\phi}\left[(1,\mathbf{\hat{n}_{\psi}})\right] \in \left[0,1\right] $ for any four-vector $(1,\mathbf{\hat{n}_{\psi}}) \leftrightarrow |\psi\rangle\langle\psi|=P_{\psi}$. In such a case,
$f_{\phi}\left[(1,\mathbf{\hat{n}_{\psi}})\right]=2c+k_0+\mathbf{\hat{n}_{\psi}}\cdot \mathbf{k}=1/2+\mathbf{\hat{n}_{\psi}}\cdot \mathbf{k} \in \left[0,1\right] $, i.e.,
\begin{equation}
-\frac{1}{2} \leq |\mathbf{k}| \cos{\theta} \leq \frac{1}{2}, \label{kk}
\end{equation}
where $\cos{\theta}=\mathbf{\hat{n}_{\psi}}\cdot \mathbf{\hat{k}} $ spans the interval $ \left[-1,1\right]$ under variation of $\mathbf{\hat{n}_{\psi}}$. This implies that $|\mathbf{k}|=1/2$, hence $\mathbf{k}=\mathbf{\hat{n}_{\phi}}/2$, and we can finally write
\begin{equation}\label{gg}
f_{\phi}\left[(1,\mathbf{\hat{n}_{\psi}})\right]=\frac{1}{2}\left(1+\mathbf{\hat{n}_{\phi}}\cdot \mathbf{\hat{n}_{\psi}}\right). 
\end{equation}

Using $P_{\psi }=| \psi \rangle \langle \psi
| =\left( \openone+\mathbf{\hat{n}}_{\psi }\cdot
\boldsymbol{\sigma }\right) /2$ and similarly for $P_{\phi
}=| \phi \rangle \langle \phi | $,
we can write $f_{\phi }(P_{\psi })$ in the standard form
\begin{equation}
f_{\phi }(P_{\psi })
=| \langle \phi |\psi \rangle |%
^{2}=\Tr \left( P_{\phi }P_{\psi }\right). \label{final}
\end{equation}

The measure $f_{\phi}$ we have obtained under the above requirements can be consistently interpreted as a probability measure. We have put our requirements on a function $f_{\phi}$ that applies to vectors $r\in V_4$ in general. It is just in order to fix some of the parameters that define $f_{\phi}$, i.e., $c$ and $k=(k_0,\mathbf{k})$, see Theorem \ref{t1}, that we conveniently applied $f_{\phi}$ to some vectors in $V_4$ having the particular form $(1,\mathbf{\hat{n}})$. These vectors belong to $V_4$ in spite of carrying only two independent parameters. Now, as for the function $f_{\phi}$, it has not been completely fixed. Though we know its action on vectors of the form $(1,\mathbf{\hat{n}})$, see Eq.~(8), we do not know its action on more general vectors $r\in V_4$. This is because we have fixed only $\mathbf{k}=\mathbf{\hat{n}}_{\phi}/2$, while $c$ and $k_0$ remain yet undetermined. In order to fix them, we can consider the vector $(-1,\mathbf{\hat{n}_{\phi}})$, which is orthogonal to $|\phi\rangle \leftrightarrow r_{\phi}\equiv (1,\mathbf{\hat{n}_{\phi}})$. Thus, we must consistently require that
\begin{equation}\label{44}
  f_{\phi}\left[(-1,\mathbf{\hat{n}_{\phi}})\right]= 2c-k_{0}+\mathbf{\hat{n}_{\phi}}\cdot \mathbf{k}=2c-k_{0}+\frac{1}{2} = 0.
\end{equation}
On account of the above equation and $2c+k_0=1/2$, we get $c=0$ and $k_0=1/2$. Hence, $k=r_{\phi}/2$ and Theorem \ref{t1} establishes that $f_{\phi}$ is a linear function given by $f_{\phi}(r)=k\cdot r$, i.e.,
\begin{equation}\label{final0}
  f_{\phi}[(r_0,\mathbf{r})]=\frac{1}{2}\left( r_0+\mathbf{\hat{n}_{\phi}}\cdot \mathbf{r}\right).
\end{equation}
On view of $(r_0,\mathbf{r})\leftrightarrow R_{\psi}\equiv \rho_{\psi}=
\sum_{\mu} r_{\mu} \sigma_{\mu}/2$, see Eq.~(\ref{roy}), and $(1,\mathbf{\hat{n}}_{\phi})\leftrightarrow P_{\phi}\equiv \rho_{\phi}= \left(
\openone+\mathbf{\hat{n}}_{\phi }\cdot \boldsymbol{\sigma }\right)/2$, see Eq.~(\ref{proy}), we can also write
\begin{equation}\label{final2}
  f_{\phi}[(r_0,\mathbf{r})]= \Tr(\rho_{\phi}^{\dagger}\rho_{\psi}).
\end{equation}
In summary, $f_{\phi}(r)$ is nothing but a scalar product. It can be specified either in vector space $V_4$, where it is given by the Euclidean scalar product, or in the space of linear operators acting on $\mathcal{H}_{2}$, where it is given by the Hilbert-Schmidt inner product $\Tr(A^{\dagger}B)$. Of course, $f_{\phi}(r)$ can be negative for some $r\in V_4$. But if we restrict ourselves to apply $f_{\phi}(r)$ on vectors $(1,\mathbf{\hat{n}}_{\psi})\in V_4$, then $f_{\phi}[(1,\mathbf{\hat{n}}_{\psi})]\in  \left[0,1\right]$ and we may use $f_{\phi}$ as a probability measure. It is up to us to decide which mathematical tools we employ in order to describe our experimental observations. The probability measure $f_{\phi}$ is just one of these tools. As discussed in \cite{fdz}, it is not exclusively connected to quantum phenomena.

\section{Summary}

We have seen that Gudder's theorem leads to a twofold extension of Gleason's theorem. An extension in which, first, qubits are included within the scope of the theorem and, second, Born's probability rule arises as a special case of a scalar product. Qubits may be understood as spanning a four-dimensional, real vector space $V_4$ whose elements are of the form $(r_0,\mathbf{r})$. The function $f$ that is the subject of Gudder's theorem acts on this space. It is assumed to be continuous and orthogonally additive. When dealing with vectors of the form $(1,\mathbf{\hat{n}})$, we can impose some additional requirements on $f$, which let us interpret it as a probability measure $f_{\phi}$ that is defined in terms of some fixed state $(1,\mathbf{\hat{n}}_{\phi})$. When $f_{\phi}$ acts on more general vectors $(r_0,\mathbf{r})$, then it acts as an inner product. As pointed out in \cite{fdz}, having discussed the two-dimensional vector space, we have essentially discussed all higher-dimensional vector spaces, at least with respect to Gleason's theorem and its extension.


\begin{thebibliography}{99}

\bibitem{Hall} M. J. W. Hall, arXiv: 1611.00613v2 [quant-ph]

\bibitem{Gleason} A. M. Gleason, \textit{J. Math. Mech.} \textbf{6} 885 (1957)

\bibitem{fdz} F. De Zela, \textit{Found. Phys} \textbf{46} 1293 (2016)

\bibitem{Gudder} S. P. Gudder, \textit{Stochastic Methods in Quantum Mechanics} (New York: North-Holland, 1979)

\bibitem{Busch} P. Busch, \textit{Phys. Rev. Lett.} \textbf{91} 120403 (2003)










\end{thebibliography}
\end{document}